\documentstyle[axodraw,aps,preprint]{revtex}
\title{One- and two-photon resonant spectroscopy of hydrogen and 
anti-hydrogen atoms in external electric fields.}
\author{
L. Labzowsky$^{1, 2, 3}$, D. Solovyev$^{1, 3}$, V. Sharipov$^1$, 
G. Plunien$^3$ and G. Soff\,$^3$\\ 
${}^1$ St. Petersburg State University, 198504, Petrodvorets, 
St. Petersburg, Russia\\ 
${}^2$ Petersburg Nuclear Physics Institute, 188350, Gatchina, 
St. Petersburg, Russia\\ 
${}^3$ Technische Universit$\ddot{a}$t Dresden, Mommsenstrasse 13, 
D-01062, Dresden, Germany}
\begin{document}
\maketitle
\begin{abstract}
The resonant spectra of hydrogen and anti-hydrogen atoms in the presence of 
an external electric field are compared theoretically. It is shown that 
nonresonant corrections to the transition frequency contain terms linear in 
the electric field. The existence of these terms does not violate space and 
time parity and leads to a difference in the resonant spectroscopic 
measurements for hydrogen and anti-hydrogen atoms in an external electric 
field. The one-photon $1s-2p$ and the two-photon $1s-2s$ resonances are 
investigated.
\end{abstract} 

\vspace{30pt}
\section{Introduction}
Recent experimental success in the production of anti-hydrogen 
atoms \cite{1}, \cite{2} makes proposals for the 
search of CPT-violating effects via accurate comparison between spectra 
of hydrogen (H) and anti-hydrogen ($\overline{{\rm H}}$) atoms \cite{3} 
rather realistic.
The most accurate experiments have been performed recently for the one-photon 
$1s-2p$ and two-photon $1s-2s$ resonances in the H atom \cite{4}-\cite{6}.

The situation necessitates the investigation of all possible reasons for 
differences in the H and $\overline{{\rm H}}$ spectra. One of these reasons 
is due to the presence of the external electric field. The electric field is 
included in the scheme of the two-photon resonance measurement \cite{5}, 
\cite{6} and can also be present in other  facilities, where experiments with 
antiprotons are intended to be performed.
It is well known that transition rates in atoms can depend linearly on the 
external electric field \cite{7}-\cite{9} and thus may differ for the H and 
$\overline{{\rm H}}$ atoms, respectively.

The energy levels of atoms can not depend linearly on the electric field 
without violation of P and T (space and time) invariance. The known "linear" 
Stark effect in a H atom provides a dependence of atomic energy levels on the 
absolute value ${\cal E}=|\vec{\cal E}|$ but not on the electric field 
strength $\vec{\cal E}$ and thus will not lead to differences in H and 
$\overline{{\rm H}}$ spectra. However, there is a limit for unambigous 
measurements of the transition frequency, i.e. for the determination of the 
energy levels set by the distortion (asymmetry) of the natural spectral line 
profiles. This asymmetry arises mainly due to the nonresonant (NR) 
corrections to the resonant Lorentz line profile. Nonresonant corrections 
have been first introduced by F. Low \cite{10} and recently evaluated for 
$1s-2p$ \cite{11}, \cite{12} and estimated for $1s-2s$ \cite{12}, \cite{13} 
transitions. Another source of the line profile asymmetry is the dependence 
of the level width $\Gamma(\omega)$ on the frequency $\omega$ \cite{13}.

The asymmetric line profile cannot be described by the two parameters energy 
$E$ and width $\Gamma$, respectively. Thus the determination of the 
transition frequency becomes ambigous and knowledge about 
the nonresonant  corrections is 
required in order to compare the results of a transition-frequency 
measurement (by the deducing the maximum of the frequency distribution or the 
line center etc) with the theoretical value for the energy difference between 
the two levels. The nonresonant corrections, unlike all other energy corrections 
depend on the excitation mechanism.

Moreover, in the external electric field the 
nonresonant corrections can depend 
linearly on the field thus giving rise to a difference in measurements of 
frequencies in H and $\overline{{\rm H}}$ spectra. In this paper we present 
the evaluation of such an effect for the one-photon $1s-2p$  and two-photon 
$1s-2s$ resonances in H and $\overline{{\rm H}}$ atoms, respectively.

\section{Nonresonant corrections to the \boldmath{$1s-2p$} one-photon resonance}

Consider the process of elastic resonant photon scattering on the ground 
state of H and $\overline{{\rm H}}$ atoms in the presence of an external 
electric field. The photon frequency may be close to the $1s-2p_{1/2}$ 
transition energy. The external electric field is assumed to mix $2p_{1/2}$ 
and $2s_{1/2}$ states, but not the $2p_{3/2}$- and $2s_{1/2}$-states. The 
major contribution to the effect under considerating originates from the 
$2s_{1/2}$ state which is closest to the resonant $2p_{1/2}$ state. 
Accordingly, the magnitude ${\cal E}$ of the electric field optimal for 
observing the effect corresponds to a Stark parameter $\xi\approx 1$. This 
parameter is defined as the ratio $\xi=S/\Delta E_L$ between the Stark matrix 
element $S=e |<2s|\vec{d}|2p>| {\cal E}$, with $\vec{d}$ being the electric 
dipole moment operator for the atomic electron and the corresponding Lamb 
shift $\Delta E_L=E_{2s}-E_{2p}$, $e$ is the electron  charge. 
Employing atomic units one obtains $S=\mp\sqrt{3}{\cal E}$ in the case of H and 
$\overline{{\rm H}}$ atoms (the signs $\mp$ correspond to H and  
$\overline{{\rm H}}$, respectively).

The case with $\xi\ll 1$ was considered earlier in \cite{14}.
For electric fields with $\xi>1$  the former $2s$, $2p_{1/2}$ levels begin to 
repel each other and thus diminish the effect. The value $\xi =1$ 
corresponds to the field strength 475 V/cm \cite{15}.

The process considered is described by the Feynman graphs depicted in Fig. 1. 
Part a) of this figure corresponds to the resonant term while part b) 
represents the dominant contribution of the  nonresonant correction. 
The contribution 
of other intermediate states will be neglected.

Within this approximation the total transition amplitude of the process can 
be expressed as 
\begin{eqnarray}
\label{1}
A\sim\frac{A_{2p'_{1/2}}(\vec{k'}\vec{e'})
A^*_{2p'_{1/2}}(\vec{k}\vec{e})}{x-\frac{i}{2}\Gamma_{2p'_{1/2}}}+
\frac{A_{2s'_{1/2}}(\vec{k'}\vec{e'})
A^*_{2s'_{1/2}}(\vec{k}\vec{e})}{x+\Delta E_L}.
\end{eqnarray}
Here $A_{2p'_{1/2}}(\vec{k'}\vec{e'})$ and $A_{2s'_{1/2}}(\vec{k'}\vec{e'})$ 
represent the individual transition amplitudes for the photon absorption 
processes $1s\rightarrow 2p'_{1/2}$ and $1s\rightarrow 2s'_{1/2}$, 
respectively, and $\vec{k'}, \vec{e'}$ denote the momentum and the 
polarization vector of the incident photon. Similarly, 
$A^*_{2p'_{1/2}}(\vec{k}\vec{e})$ and $A^*_{2s'_{1/2}}(\vec{k}\vec{e})$ 
denote the corresponding amplitudes for the emission processes 
$2p'_{1/2}\rightarrow 1s$ and $2s'\rightarrow 1s$, respectively, and 
$\vec{k}, \vec{e}$ are the momentum and the polarization vector of the 
emitted photon. Furthermore, we have introduced the energy parameter 
$x\equiv E_{2p'_{1/2}}-E_{1s}-\omega$ and the total width 
$\Gamma_{2p'_{1/2}}$ of the level $2p'_{1/2}$ in the denominators. By 
$2p'_{1/2}$ and $2s'$ we denote the states in the electric field, that go 
over to $2p_{1/2}$ and $2s$ in the absence of this field. We assume that 
under the influence of the external electric field the conditions 
$\Gamma_{2s'}\le \Gamma_{2p'_{1/2}}$ and $\Gamma_{2p'_{1/2}}\ll \Delta E_L$ 
hold. Accordingly, we can omit the width $\Gamma_{2s'}$ in the second 
denominator in Eq. (1).

Summation over the polarizations of the incident and emitted photons leads to 
an expression for the differential cross section
\begin{eqnarray}
\label{2}
\sigma\sim\frac{W^a_{2p'_{1/2}}(\vec{k'})
W^e_{2p'_{1/2}}(\vec{k})}{x^2+\frac{1}{4}\Gamma^2_{2p'_{1/2}}}+
\frac{W^a_{2s'_{1/2}}(\vec{k'})W^e_{2s'_{1/2}}(\vec{k})}{(x+\Delta E_L)^2}.
\end{eqnarray}
Here $W^a_{2p'_{1/2}}(\vec{k'})$ and $W^a_{2s'_{1/2}}(\vec{k'})$ denote the 
differential absorption probabilities and $W^e_{2p'_{1/2}}(\vec{k}), 
W^e_{2s'_{1/2}}(\vec{k})$ are the differential emission probabilities for the 
transitions $2p'_{1/2}\leftrightarrow 1s$ and $2s'_{1/2}\leftrightarrow 1s$, 
respectively.

Even for $\xi=1$ the total width $\Gamma_{2p'}$ is only weakly affected by 
the electric field. Therefore we can set 
$\Gamma_{2p'_{1/2}}\equiv\Gamma_{2p}$.

The first term in Eq. (2) represents the usual resonance expression 
(Lorentz line profile) while the second term provides the 
nonresonant correction. To 
simplify the evaluations we suppose that the resonant transition frequency 
$\omega_{{\rm res}}$ is defined by the maximum of the cross section, i.e. from the 
condition $d\sigma(x)/dx=0$. If we neglect the nonresonant correction, 
$\omega_{{\rm res}}$ 
corresponds to $x=0$. Inclusion of the nonresonant correction yields
\begin{eqnarray}
\label{3}
x_{{\rm NR}}=-\frac{1}{16}\frac{\Gamma^4_{2p}}{\Delta E_L^3}
\frac{W^a_{2s'}(\vec{k'})W^e_{2s'}(\vec{k})}{W^a_{2p'}(\vec{k'})
W^e_{2p'}(\vec{k})}.
\end{eqnarray}

In the absence of the electric field $(\xi=0)$ the nonresonant
correction (3) is 
vanishingly small. In this case $\Gamma_{2p}=0.04\alpha^3, 
\Delta E_L=0.4\alpha^3, W^a_{2p}(\vec{k'})\approx W^e_{2p}(\vec{k})
\approx\Gamma_{2p}$ and $W^a_{2s}(\vec{k'})\approx W^e_{2s}(\vec{k})
\approx\Gamma_{2s}=10^{-3}\alpha^9$, respectively (in atomic units). Then one 
obtains $x_{{\rm NR}}\approx -10^{-22}$ Hz. The main contribution to the shift of 
the maximum $x_{{\rm NR}}$ in the absence of the field originates from the 
interference terms between the $1s-2p_{1/2}$ and $1s-np_{1/2}$ transitions 
and from the quadratic $1s-2p_{3/2}$ NR correction \cite{11}-\cite{13}. This 
yields $x_{{\rm NR}}=-7.7$ Hz. However, if the electric field is present ($\xi=1$) 
the correction (3) becomes by far the largest: $x_{{\rm NR}}\approx-6.25$ kHz.

\section{The nonresonant correction in external electric fields}
We are interested in the contributions to  Eq. (3) linear in the electric field. 
The differential emission probability for the transition $2s'\rightarrow 1s$ 
has been evaluated in \cite{7}-\cite{9} and takes the form
\begin{eqnarray}
\label{4}
W^e_{2s'}(\vec{k})=W_{2s}-\xi\frac{\Gamma_{2p}}{\Delta E_L}(\vec{\nu}\vec{F})
(W^e_{2s}W^e_{2p})^{1/2}+\xi^2W^e_{2p}.
\end{eqnarray}
Here we introduce the notations $\vec{\nu}=\vec{k}/\omega$ and 
$\vec{F}=\vec{{\cal E}}/|\vec{\cal E}|$. In Eq. (4) the approximation 
$\Gamma_{2p}\ll\Delta E_L$ is employed again.

The term linear in the electric field leads to a difference in the transition 
probabilities for H and $\overline{{\rm H}}$ atoms in the external electric 
field. For $\xi=1$ this difference is about 
$W^e_{2p}\approx\Gamma_{2p}(1\mp 10^{-7})$, where the different signs 
$\mp$ correspond to the hydrogen and anti-hydrogen atoms, respectively.

A similar expression can be derived for $W^e_{2p'}$:
\begin{eqnarray}
\label{5}
W^e_{2p'}(\vec{k})=W_{2p}+\xi\frac{\Gamma_{2p}}{\Delta E_L}(\vec{\nu}\vec{F})
(W^e_{2s}W^e_{2p})^{1/2}+\xi^2W^e_{2s}
\end{eqnarray}
The most important difference between Eqs. (4) and (5) is that in Eq. (4) the 
leading term is the last one (provided the external field is not too small),
while in Eq. (5) the major contribution comes 
from the first term. Similar 
expressions for the corresponding absorption probabilities can be derived.

Suppose that we are interested in the correlation $(\vec{\nu}\vec{F})$ 
between the direction of the photon emission and the electric field. 
Accordingly, we have to set $W^a_{2s'}(\vec{k'})=\xi^2W^a_{2p}, 
W^a_{2p'}(\vec{k'})=W^a_{2p}$, and to use Eqs. (4) and (5) for 
$W^e_{2s'}(\vec{k}), W^e_{2p'}(\vec{k})$, omitting there the first and the 
last terms correspondingly. Expansion of the denominator in Eq. (3) yields
\begin{eqnarray}
\label{6}
x_{{\rm NR}}=-\frac{1}{16}\frac{\Gamma_{2p}^2}{\Delta E_L^3}\xi^4\left[1-
\left(\xi+\frac{1}{\xi}\right)\frac{\Gamma_{2p}}{\Delta E_L}
\left(\frac{W^e_{2s}}{W^e_{2p}}\right)^{1/2}(\vec{\nu}\vec{F})\right].
\end{eqnarray}
For $\xi=1$ the nonresonant
correction linear in the electric field is given by
\begin{eqnarray}
\label{7}
\delta x_{{\rm NR}}=-\frac{1}{8}\frac{\Gamma^5_{2p}}{\Delta E^4_{L}}
\left(\frac{W^e_{2s}}{W^e_{2p}}\right)^{1/2}(\vec{\nu}\vec{F})
\end{eqnarray}
or, numerically, $\left|\delta x_{{\rm NR}}\right|\approx 10^{-4}$ Hz.

The formal T-noninvariance of the factor $(\vec{\nu}\vec{F})$ in Eq. (7) 
($\vec{\nu}$ and $\vec{F}$ are T-odd and T-even vectors, respectively) is 
compensated by the linear dependence on $\Gamma_{2p}$ in Eq. (7): this is a 
well known imitation of T-nonivariance in unstable systems as it has been 
predicted by Zel$'$dovich \cite{16}. Such a case of T-noninvariance has been 
discussed in \cite{7}-\cite{9}.

 Eq. (7) defines a measurable difference between the  $1s-2p$ transition 
 frequency in H and $\overline{{\rm H}}$ atoms. A direct observation of this 
 difference seems rather unlikely at present, since the inaccuaracy of the 
 Lyman-alpha transition measurement is about 6 MHz \cite{4}, i.e. more than 
 $10^{10}$ orders of magnitude larger than the correction (7).

Note, that the part of the correction $x_{{\rm NR}}$ independent on the field 
direction, and thus equal in H and $\overline{{\rm H}}$, is only  $10^3$ 
times smaller than the inaccuracy in \cite{4}. A direct observation of the 
distortion of the natural line profile due to NR corrections would be of 
special interest irrelevant of the comparison of the H and 
$\overline{{\rm H}}$ spectra.

\section{The \boldmath{$1s-2s$} two-photon resonance}

In this section we will investigate the NR corrections to the transition 
frequency in the $1s-2s$ two-photon resonance in H and $\overline{{\rm H}}$ 
atoms. In the highly accurate experiment \cite{5}, \cite{6} on the $1s-2s$ 
two-photon excitation in hydrogen an inaccuracy of 46 Hz has been achieved.

However, the experiment \cite{5}, \cite{6} is based on special time-delay 
techniques. The region, where the excitation process $1s+2\gamma\rightarrow 
2s$ occurs, is separated from the detection region by some finite distance 
$S$. In the detection region a ``small'' quenching field is applied and the 
one-photon decay $2s'\rightarrow 1s+\gamma$ is observed. With a spatial 
separation $S\approx 13$ cm and for typical velocities of the hydrogen atom 
of about $\upsilon\approx 10^{4}$ cm/sec adopted in \cite{5}, \cite{6} the 
time delay $t_D$ in the decay registration appears to be $t_D\approx 10^{-3}$ 
s. This corresponds to a time-delayed experimental width of about 1 kHz. If 
we adopt for the estimates a ``small'' electric field 
$\xi=0.1$ $({\cal E}=45.5$ V/cm$)$ the decay time $t_d$ according to Eq. (4) 
will be defined by $t_d=\left(\xi^2\Gamma_{2p}\right)^{-1}\approx 10^{-7} 
s\ll t_D$. Thus the atoms that reach the detection region will decay 
immediatly and the experimental decay line profile will be determinated 
exclusively by $t_D$.

In the absence of an external electric field the 
nonresonant corrections for $1s-2s$ 
two-photon resonance have been estimated in \cite{12}. These corrections are 
defined by the interference between the amplitudes corresponding to the 
Feynman graph of Fig. 2 with $n=2$ and the Feynman graphs in Fig. 2 with 
$n=3, 4, ...$ In this case the nonresonant correction appears to be negligible: 
$|x_{{\rm NR}}|\sim 10^{-14}$ Hz \cite{12}.

In the presence of the electric field the situation may change drastically. 
However, an exact quantum electrodynamical description of the resonant 
process with the time-delayed decay is not obvious: it would require the 
introduction of the S-matrix defined for finite time intervals. This would 
cause many difficulties, among which the problem of the renormalizability of 
the amplitudes seems to be most  striking one.

In \cite{5} a simplified quantum mechanical approach based on the density 
matrix formalism has been employed for the description of the resonance line 
shape of the $1s-2s$ two-photon excitation process with delayed registration 
by the electric field quenching. Still it is not quite evident how to 
introduce nonresonant corrections within this approach. In \cite{13} an 
estimate for $x_{{\rm NR}}$ based on purely phenomenological considerations has 
been provided. An accurate and reliable value for $x_{{\rm NR}}$ for the 
experiments \cite{5}, \cite{6} is still not available. Leaving a solution of 
this problem for future studies, in this paper we will investigate a process 
opposite to one considered in \cite{12}: the two-photon $1s-2s$ excitation 
and decay in external electric fields. The case considered here will 
correspond to $\xi=1$, while the case of Ref.\cite{12} corresponds to 
$\xi=0$.

We have to stress that our case differs also from the real experiment 
\cite{5}, \cite{6} where the electric field is present only in the region 
where the decay process takes place (``decay part''), not in the 
``excitation part''. The process with the electric field present all the 
time loses, of course, most of the advantages of the experimental approach 
developed in \cite{5}, \cite{6} and should provide much poorer accuracy. 
Still it is preferable for the search of the difference in the transition 
frequency measurements in H and $\overline{{\rm H}}$ atoms in electric 
fields.

The Feynman graphs, corresponding to the resonant two-photon excitation 
process $1s'+2\gamma\rightarrow 2s'$ and the dominant nonresonant 
contribution are depicted in Fig. 3. We neglect the correction due to the 
quadratic Stark effect for the 1s-ground state. Even for $\xi=1$ this 
correction is of relative order $10^{-12}$. Within the approximation depicted 
in Fig. 3 an expression for the differential cross 
section similar to Eq. (2) can be derived
\begin{eqnarray}
\label{8}
\sigma\sim\frac{W^a_{2s', 2\gamma}(\vec{k'})W^e_{2s', 1\gamma}
(\vec{k})}{x^2+\frac{1}{4}\Gamma^2_{2s'}}+\frac{W^a_{2p', 2\gamma}(\vec{k'})
W^e_{2p', 1\gamma}(\vec{k})}{(x+\Delta E_L)^2}.
\end{eqnarray}
The indices $2\gamma, 1\gamma$ denote the two- or one-photon processes 
$2s'\leftrightarrow 1s+2\gamma, 2p'\leftrightarrow 1s+2\gamma, 
2s'\leftrightarrow 1s+\gamma$ and $2p'\leftrightarrow 1s+\gamma$. Here we 
define $x=E_{2s'}-E_{1s}-\omega$.

Repeating the considerations presented in Section 2, we obtain from Eq. (8) 
an expression for the nonresonant  correction:
\begin{eqnarray}
\label{9}
x_{{\rm NR}}=-\frac{1}{16}\frac{\Gamma^4_{2s'}}{\Delta E_L^3}\frac{W^a_{2p', 
2\gamma}(\vec{k'})W^e_{2p', 1\gamma}(\vec{k})}{W^a_{2s', 2\gamma}(\vec{k'})
W^e_{2s', 1\gamma}(\vec{k})}
\end{eqnarray}

For simplicity, we consider only the case $\xi=1$ corresponding to the 
maximum nonresonant contribution. 
Accordingly, one obtains $\Gamma_{2s'}=\Gamma_{2p}$ 
and $W^a_{2p', 2\gamma}(\vec{k'})\approx W^a_{2s', 2\gamma}(\vec{k'}), 
W^e_{2p', 1\gamma}(\vec{k})\approx W^e_{2s', 1\gamma}(\vec{k})\approx
\Gamma_{2p}$ which yields the same  result as for the one-photon $1s-2p$ 
transition: the correction, independent from the field direction is $x_{{\rm NR}}
\approx -6.25$ kHz. A similar result arises for the correction, which depends 
on the field direction: $\left|\delta x_{{\rm NR}}\right|\approx 10^{-4}$ Hz.

\section{Conclusions}
In this paper we have shown, that resonant atomic spectra in the external 
electric fields are different for H and $\overline{{\rm H}}$ atoms, 
respectively. At first, the transition rates for the Lyman-alpha transition 
in H and $\overline{{\rm H}}$ atoms differ by 0.1 ppm for the ``optimal'' 
value of the electric field $\xi=1$. Secondly, the positions of the maxima 
of the natural line profiles for the Lyman-alpha resonances in H and 
$\overline{{\rm H}}$ atoms are shifted relative to each other by the value 
$2\cdot 10^{-4}$ Hz. The same concerns the two-photon resonance shifts in 
the process $1s+2\gamma\rightarrow 2s$ taking place in external electric 
fields. 

In an electric field with $\xi=1$ the nonresonant corrections, independent from the 
field direction appear to be 1000 times larger, than the nonresonant
 corrections in 
the absence of the field.

This enhancement is connected with the overlap of resonances with identical 
quantum numbers. This overlap has been studied theoretically for 
highly-charged ions in \cite{17}. In the case of H and $\overline{{\rm H}}$ 
atoms the $2p', 2s'$ levels in the field $\xi=1$ actually do not overlap, 
but unlike the $2p, 2s$ levels they have identical quantum numbers. Since 
these levels are still close to each other, their mutual influence results 
in large NR corrections.

We also would like to mention another effect that looks different in 
hydrogen and 
anti-hydrogen atoms placed in external electric fields. This is the effect 
of quantum beats in the Lyman-alpha radiation. It has been investigated 
theoretically and experimentally by many authors: see \cite{18}-\cite{23}. 
Quantum beats arise firstly, due to the interference of the $2s', 2p'$ 
states in the external field. However, this field-induced effect is 
independent of the field direction, i.e. it will be the same for H and 
$\overline{{\rm H}}$ atoms. If the additional requirement of a coherent 
excitation of the $2s', 2p'$ states is satisfied, the quantum-beats signal 
will be linear in the electric field \cite{22}. This implies that quantum 
beats signal for H and $\overline{{\rm H}}$ atoms after subtraction of the 
part independent on the field direction will exhibit a relative phase shift 
of $\pi$ in the same electric field. The coherent excitation may arise in 
beam-foil experiments \cite{20}-\cite{23} or with the laser excitation if 
the laser bandwidth is larger than $\Delta E_L$. In the experiment \cite{4} 
the laser bandwidth was about 10 MHz, i.e. about 100 times smaller than 
$\Delta E_L$. Using another sources of radiation with larger bandwidth one 
may hope to observe this phase shift for quantum beats in H and 
$\overline{{\rm H}}$ atoms.

\section*{Acknowledgements}

The authors are grateful to P. J. Mohr for valuable discussion. L. L. and 
D. S. acknowledge the support by RFBR grants 02-02-16579 and 02-02-06689-mas 
and by Minobrazobavanie grant E00-3.1-7. The work of L. L. was partially 
supported by NSF through the grant to ITAMP at Harvard University that 
L. L. visited in 2002. The work was completed during the stay of L. L. and 
D. S. at the Dresden University of Technology. This stay was sponsored by 
DFG and DAAD. G. P. and G. S. acknowledge financial support from BMBF, DFG 
and GSI.

\newpage
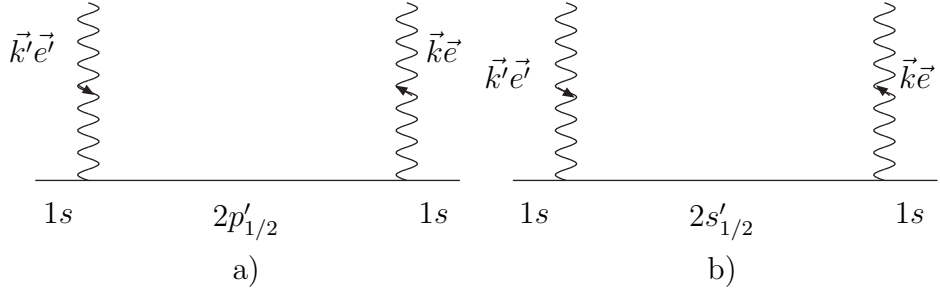
\begin{figure}
\begin{center}
\begin{picture}(450,300)(0,0)
\Line(51,128)(211,128) 
\Photon(71,128)(71,195){4}{8} \LongArrow(67,164)(72,161)\Text(50,185)[t]{$\vec{k'}\vec{e'}$}
\Photon(191,128)(191,195){4}{8} \LongArrow(193,160)(188,163)\Text(205,185)[t]{$\vec{k}\vec{e}$}
\Text(60,120)[t]{$1s$} \Text(202,120)[t]{$1s$} \Text(131,120)[t]{$2p'_{1/2}$}
\Text(131,100)[t]{a)}

\Line(231,128)(391,128) 
\Photon(251,128)(251,195){4}{8} \LongArrow(248,163)(253,160)\Text(230,175)[t]{$\vec{k'}\vec{e'}$}
\Photon(371,128)(371,195){4}{8} \LongArrow(373,160)(369,163)\Text(384,175)[t]{$\vec{k}\vec{e}$}
\Text(240,120)[t]{$1s$} \Text(382,120)[t]{$1s$} \Text(311,120)[t]{$2s'_{1/2}$}
\Text(311,100)[t]{b)}
\end{picture}
\end{center}
\caption{ 
Feynman graphs describing the elastic resonant photon scattering on the 
ground state of hydrogen or anti-hydrogen atoms. The solid lines correspond 
to bound atomic electrons (positrons), the wavy lines with arrows correspond 
to the incident and emitted photons, respectively. The indices 
$2p'_{1/2}, 2s'_{1/2}$ denote the atomic states in the electric field. 
In the absence of the field these levels go over to the 
$2p_{1/2}, 2s_{1/2}$ states. The ground state 1s is assumed to be affected 
by this field. The graph a) presents the resonant term and the graph b) 
presents the dominant nonresonant correction.
}
\end{figure}

\newpage
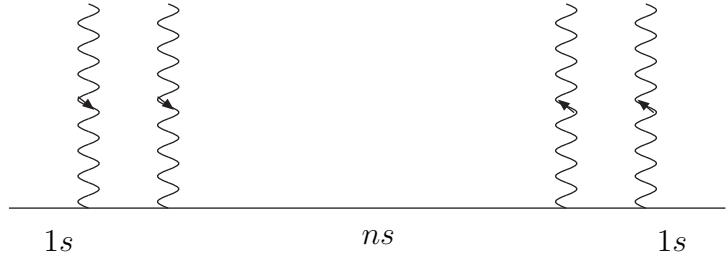
\begin{figure}
\begin{center}
\begin{picture}(350,300)(0,0)
\Line(61,128)(331,128) 
\Photon(91,128)(91,205){4}{8} \LongArrow(87,170)(92,166)
\Photon(121,128)(121,205){4}{8}\LongArrow(117,170)(122,166)
\Photon(301,128)(301,205){4}{8} \LongArrow(304,164)(299,168)
\Photon(271,128)(271,205){4}{8}\LongArrow(274,164)(269,168)
\Text(80,120)[t]{$1s$} \Text(312,120)[t]{$1s$} \Text(201,120)[t]{$ns$}
\end{picture}
\end{center}
\caption{
The Feynman graph that describes the two-photon excitation process 
$1s+2\gamma\rightarrow 2s$ in the absence of an electric field. The resonant 
term corresponds to $n=2$, the nonresonant terms correspond to 
$n=3, 4, ...$ The notations are the same as in Fig. 1.
}
\end{figure}

\newpage
\begin{figure}
\begin{center}
\begin{picture}(450,300)(0,0)

\Line(51,128)(211,128) 
\Line(71, 126)(191, 126)
\Photon(71,128)(71,195){4}{8} \LongArrow(67,164)(72,161)
\Photon(91,128)(91,195){4}{8} \LongArrow(87,164)(92,161)
\Photon(191,128)(191,195){4}{8} \LongArrow(193,160)(188,163)
\Text(60,120)[t]{$1s$} \Text(202,120)[t]{$1s$} \Text(131,120)[t]{$2s'_{1/2}$}
\Text(131,100)[t]{a)}

\Line(231,128)(391,128) 
\Line(251, 126)(371, 126)
\Photon(251,128)(251,195){4}{8} \LongArrow(248,163)(253,160)
\Photon(271,128)(271,195){4}{8} \LongArrow(268,163)(273,160)
\Photon(371,128)(371,195){4}{8} \LongArrow(373,160)(369,163)
\Text(240,120)[t]{$1s$} \Text(382,120)[t]{$1s$} \Text(311,120)[t]{$2p'_{1/2}$}
\Text(311,100)[t]{b)}
\end{picture}
\end{center}
\caption{
The Feynman graphs corresponding to the two-photon resonant excitation of 
H and  $\overline{{\rm H}}$ in an external electric field. The double solid 
lines denote bounded atomic electrons (positrons) in an external electric 
field. The ground state is assumed to be unaffected by the electric field. 
The other notations are the same as Figs. 1 and 2. Fig. 3a) corresponds to 
the resonant contribution and Fig. 3b) corresponds to the leading nonresonant 
contribution, respectively.
}
\end{figure}
\end{document}